\documentclass[a4paper,fleqn,usenatbib]{mnras}

\usepackage[T1]{fontenc}
\usepackage{ae,aecompl}
\usepackage{graphicx}

\usepackage{epsfig}
\usepackage{amsmath}	
\usepackage{amssymb}	

\newcommand{\mlya}{${\rm Ly\alpha}$}
\newcommand{\mlyb}{${\rm Ly\beta}$}

\def\review#1{{#1}}

\title[The IGM temperature at $z\sim 2.75$]{A new Measurement of the Intergalactic Temperature at $z\sim 2.55-2.95$}

\author[A.Rorai et al.]{
Alberto Rorai$^{1,2}$\thanks{E-mail: arorai@ast.cam.ac.uk},
Robert F. Carswell$^{2}$,
Martin G. Haehnelt$^{1,2}$,
George D. Becker$^{3}$,
\newauthor James S. Bolton$^{4}$,
Michael T. Murphy$^{5}$
\\
$^{1}$Kavli Institute for Cosmology and Institute of Astronomy, Madingley Road, Cambridge CB3 0HA, United Kingdom\\
$^{2}$Institute of Astronomy, Madingley Road, Cambridge CB3 0HA, United Kingdom\\
$^{3}$Department of Physics \& Astronomy, University of California, Riverside, 900 University Avenue, Riverside, CA,92521, USA\\
$^{4}$School of Physics and Astronomy, University of Nottingham, University Park, Nottingham, NG7 2RD, United Kingdom\\
$^{5}$Centre for Astrophysics and Supercomputing, Swinburne University of Technology, Hawthorn, Victoria 3122, Australia
}

\date{Accepted XXX. Received YYY; in original form ZZZ}

\pubyear{2017}

\begin{document}
\label{firstpage}
\pagerange{\pageref{firstpage}--\pageref{lastpage}}
\maketitle


\begin{abstract} 
We present two  measurements of the  temperature-density 
relationship (TDR) of the intergalactic medium (IGM) in the redshift range $2.55 < z < 2.95$ using a sample of 
13 high-quality quasar spectra and high resolution numerical simulations of the IGM.
Our approach is based on fitting the neutral hydrogen column density $N_{HI}$ 
and the Doppler parameter $b$ of the absorption lines in the \mlya\ forest. 
The first measurement is obtained using a novel 
Bayesian scheme which takes into account the statistical correlations between the parameters
characterising  the lower cut-off of the $b-N_{HI}$ distribution  
and the power-law parameters $T_0$ and $\gamma$  describing the TDR.
This approach yields $T_0/ 10^3\, {\rm K}=15.6 \pm 4.4 $ and $\gamma=1.45 \pm 0.17$ independent 
of the assumed pressure smoothing of the small scale density field. 
In order to explore the information contained in the overall $b-N_{HI}$ distribution  
rather than only the lower cut-off,  we obtain a second measurement 
based on  a similar Bayesian analysis of the  median Doppler parameter  for  separate column-density ranges 
of the absorbers.  In this case  we obtain $T_0/ 10^3\, {\rm K}=14.6 \pm 3.7$ and
$\gamma=1.37 \pm 0.17$ in good agreement with the first measurement.  
Our Bayesian analysis  reveals strong anti-correlations 
between the inferred $T_0$ and $\gamma$ for both methods as well as  an anti-correlation 
of the  inferred $T_0$ and  the  pressure smoothing length for the second method,  suggesting that the measurement 
accuracy can in the latter  case be substantially increased if independent constraints on the smoothing are obtained. Our 
results  are in good agreement with other  recent measurements of the thermal state of  the IGM probing 
similar (over-)density ranges.
\end{abstract}




\section{Introduction} 
The intergalactic medium (IGM), containing the overwhelming 
majority of the Universe's baryons, retains key information about 
the cosmic transformations that occurred during helium and 
hydrogen reionisation  \citep[e.g.,][]{HuiGnedin97,GnedHui98,TheunsMoSchaye01,Theuns02a,HH03}.
Many authors have studied the signature of thermal heating caused by 
the ionizing photons
in the absorption profiles of the \mlya\ forest, with the goal of probing
the temperature-density relation (TDR) of the intergalactic medium (IGM) 
 at $z=2-4$ \citep[e.g.][]{Haehnelt98,Schaye1999,McDonald2001,Zald01,Lidz09}. 
These works 
have been motivated by the simple form that the low-density TDR 
should take according to theoretical predictions, and by 
the potential implications for the history of reionization. Analytical
studies and hydrodynamics simulations have indicated 
that the low-density gas 
in the IGM should be concentrated in a narrow region of the temperature-density
plane along a power law  
\begin{equation}\label{eqn:TDR}
T=T_0 \Delta^{\gamma -1},
\end{equation}
where $T_0$ is the temperature at mean density, $\Delta$ is the density
divided by the mean of the Universe and $\gamma$ is the index of the
power-law relation \citep{HuiGnedin97}. 
The shape of the TDR at different redshifts  is dependent on the 
timing of reionization, on the nature of the sources and physical
mechanisms responsible for the heating.  
If photoheating of residual neutral hydrogen is the dominant heat 
source, then it is predicted that $\gamma\approx 1.6$ well after
reionization \citep{HuiGnedin97,McQuinn2015}. 

\begin{table*}
\label{table:data}
\centering
\begin{tabular}{@{}lccccll@{}}
\hline\hline\\
Object        &  $z_{\rm em}$ & $\Delta z$& Source & S/N & pixel size & ESO program/reference \\
\hline\\
Pks2126-158	&	3.28		& 0.0939& UVES	&	40 -- 90  &  2.5 km\,s$^{-1}$ &166.A-0106(A)\\
Q0347-383	&	3.23		& 0.1485& UVES	&	50 -- 55  &  2.5 km\,s$^{-1}$ &68.B-0115(A)\\
J134258-135559	& 3.21	& 0.2545& UVES	&	30 -- 50  &  2.5 km\,s$^{-1}$ &68.A-0492(A)\\
HS1425+6039	&	3.18		& 0.2356& HIRES	&	55 -- 75  & 0.04\,A & Sargent\\
Q0636+6801	&	3.17	& 0.3152& HIRES	&	35 -- 60  & 0.04\,A & Sargent\\
J210025-064146 &	3.14	& 0.1356& HIRES	&	20 -- 25  & 1.3 km\,s$^{-1}$ & KODIAQ \cite{Omeara2015}\\
Q0420-388	&	3.12	& 0.2884& UVES	&	75 -- 120  &  2.5 km\,s$^{-1}$&166.A-0106(A)\\
HE0940-1050	&	3.08	& 0.2910& UVES	&	35 -- 70  &  2.5 km\,s$^{-1}$&166.A-0106(A)\\
GB1759+7539	&	3.05	& 0.1859& HIRES	&	27 -- 32 & 2.0 km\,s$^{-1}$ & \cite{Outram1999}\\
J013301-400628 &	3.02	& 0.1929& UVES	&	25 -- 35 &  2.5 km\,s$^{-1}$&69.A-0613(A),073.A-0071(A),074.A-0306(A)\\
J040718-441013 &	3.02	& 0.0962& UVES	&	40 -- 115	&  2.5 km\,s$^{-1}$&68.A-0361(A),68.A-0600(A),70.A-0017(A)\\
J224708-601545	& 3.00	& 0.3293&	 UVES	&	35 -- 75 &  2.5 km\,s$^{-1}$&075A-158(B)\\
HE2347-2547	&	2.88	& 0.2205& UVES	&	95 -- 125 &  2.5 km\,s$^{-1}$&077.A-0646(A),166.A-0106(A)\\
\hline\hline
\end{tabular}
\caption{List  of spectra analysed in this work. 
Listed, from left to right, are the object name, redshift, Ly$\alpha$ absorption redshift range used,
the instrument used for observation, the signal-to-noise range (per pixel) at the
continuum level, the reduced data pixel size, and either the ESO program number (UVES) or the source of the data (HIRES).
}
\end{table*}

Although several statistical analyses of the \mlya\ forest 
find values of $\gamma$ close to this prediction \citep{Rudie2012,Bolton14,Boera2016},
other measurements which consider the \mlya\ flux probability distribution 
function (PDF), have claimed  evidence for an inverted ($\gamma < 1$) TDR
\citep{Bolton08,VielBolton09,Calura2012,Garzilli2012} for which 
unconventional heating mechanisms such as blazar heating have been invoked as an explanation
\citep{Blazar1,Blazar2,Blazar3,Puchwein12,Lamberts2015}.
Some authors have suggested that this discrepancy could be ascribable
to unaccounted effects from systematic uncertainties due to ``continuum fitting'' 
of QSO absorption spectra necessary for the calculation of the flux
PDF \citep{Lee2012} or to an overestimation
of the statistical significance of the measurements \citep{Rollinde2013}.
An alternative solution to reconcile the apparent discrepancies 
between the measurements and the  expected thermal state 
of a photo-heated IGM was proposed in \cite{Rorai2017}, who
analyzed the PDF of the low-opacity pixels in a 
very high signal-to-noise quasar 
spectra in order to constrain the TDR in the low-density IGM. 
\cite{Rorai2017} found that uncertainties in the continuum 
placement   alone cannot explain the discrepancy with conventional models for the thermal state 
of the IGM.  Instead, they found that a flat or inverse TDR (with high 
temperature in underdense regions) is indeed 
favoured by the PDF, though perhaps only at very low densities 
($\Delta \lesssim$ 1).
They also showed that different \mlya\ forest statistics 
that give discrepant results, like the power spectrum or 
those based on line-fitting methods, are sensitive to disjoint  density ranges ($\Delta \gtrsim $ 2-3).  \cite{Rorai2017} thus  
challenged  the description  of the low-density TDR as a  
single spatially-invariant power law.

To investigate  the low density TDR further, here we undertake  
a traditional Voigt-profile fitting decomposition  of  the \mlya\  
forest absorption 
in the redshift range $2.55<z<2.95$ for a sample of 13 quasar spectra.
We use a set of IGM models obtained by post-processing 
high resolution hydrodynamics simulations and
generate model spectra with the same noise and resolution characteristics. 
We apply the same Voigt decomposition to the these spectra so that they may 
then be compared directly with the telescope data.
Following \cite{Schaye1999,Schaye00,Ricotti00,Rudie2012,Bolton14}, we analyse the
shape of the cutoff for narrow lines in the plane defined by 
the column density $N_{HI}$ and the line width $b$.
Moreover, we introduce a 
Bayesian formalism to study not only the uncertainties on the 
inferred thermal parameters, including the pressure smoothing,
but also their degeneracies. We further  develop a new technique
based on the medians of the $b$ distribution for separate 
column-density ranges, in order to exploit the information contained 
in the bulk of the distribution in the $N_{HI}-b$ plane. 

This article is structured as follows. We start by presenting in
\S~\ref{Sec:data} the sample of quasar spectra we use in our analysis
and how these data are treated, in particular with respect to metal 
contaminants. We then describe the hydrodynamics simulations and 
the models to which we compare the data (\S~\ref{Sec:hydro}). In
\S~\ref{Sec:method} we explain how we analyse the statistical 
properties of \mlya\ lines in the forest to extract the information
about the thermal properties of the IGM. The results of this analyses 
are illustrated in \S~\ref{Sec:results} and subsequently discussed 
in \S~\ref{Sec:discussion}, where we also examine agreements and 
disagreements with previous studies. We draw our final conclusions 
in \S~\ref{Sec:conclusions}.

\section{data}\label{Sec:data}

\begin{figure*}
\includegraphics[width=\textwidth]{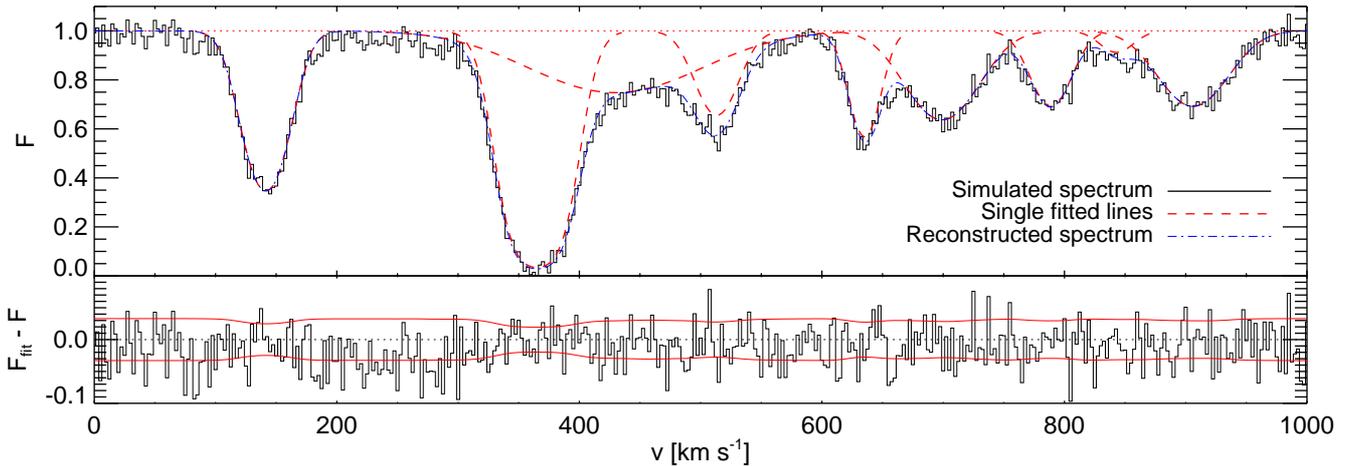}
\caption{\label{Fig:vpfit} 
Decomposition of the \mlya\ forest in individual absorption lines. 
\review{\emph{Top panel: }}the black histogram represents the \mlya\ absorption in a sight line 
from one of our models shown in velocity space. The average signal-to-noise
level of this spectrum is 33.2. The red dashed 
curves are the individual fitted absorbers, while the total reconstructed 
absorption is shown as the  blue
dot-dashed curve. The red dotted line follows the continuum ($F=1$). 
\review{\emph{Bottom panel:} the deviation of the reconstructed 
flux $F_{\rm fit}$ from the original spectrum $F$, compared with the 
assumed noise level for this synthetic spectrum (delimited by the
red solid lines).}
Overall, the combined fit is a good approximation to 
the original spectrum, except in regions characterized
by flat absorption profiles (for example around $v=200-300 $ km s$^{-1}$ )
which cannot be decomposed into individual lines. Note also that rather broad 
lines are sometimes required  to reproduce the absorption 
profile, which are not obviously interpretable as single 
absorbers ({\it e.g.} in  the case of the line centred around $v\sim 450$
km s$^-1$). The parameters of these  broad lines are also particularly sensitive to 
systematic uncertainties due to the continuum placement.}
\end{figure*}

A sample of high signal-to-noise ratio quasar spectra from
ESO Ultraviolet and Visual Echelle Spectrograph
 \citep[UVES]{UVES} and Keck 
high resolution echelle spectrograph \citep[HIRES]{HIRES} archival 
data with coverage
of the \mlya\ forest in the redshift range $2.55 < z < 2.95$ was
selected. \review{This redshift range is chosen to complement the
reanalysis by \cite{Bolton14} of the data presented by
\cite{Rudie2012}, which provided constraints on the TDR
parameters at lower redshift. Note that the methods illustrated in 
this paper can in principle be applied to higher redshift data, 
but we found that at $z > 3.2$ the stronger blending of \mlya\ lines
makes the decomposition into Voigt profiles increasingly ambiguous.} 
To have a large segment of the chosen absorption
redshift range clear of the quasar proximity region (chosen to
be within 4000 km s$^{-1}$
of the quasar redshift) then requires
that the emission redshift $z_{\rm em} \gtrsim 2.85$, 
and the need to avoid
\mlyb\ blending with the \mlya\ forest leads to $z_{\rm em} \lesssim 3.30$. 
A list
of the 13 objects used, and the characteristics of the reduced
spectra, is given in Table~\ref{table:data}.
The exposures for the nine 
UVES spectra were reduced using the European Southern 
Observatory (ESO) UVES Common Pipeline Language software 
(v4.2.8) and combined using UVES\_popler \citep{MurphyPopler},
as described in detail in \cite{Boera2014}. 
The two spectra kindly provided by W.L.W. Sargent were reduced
using {\sc makee} \citep{Makee}. Continuum estimates
in the forest are based on fitting low order curves to the high
points in the spectra.

Here we wish to compare the results of fitting Voigt
profiles to the \mlya\ forest using {\sc VPFIT}
\citep{vpfit} to the observational spectra with
those from simulated data (see \S~\ref{Sec:hydro}), 
so we have to be aware of some
features in the data which cannot be reproduced in the simulations,
and some restrictions the simulations may place on the way
the observations are analyzed. These are:
\begin{itemize}
\item The resolution of the object spectra is not accurately
known, since the observations were not always slit limited
and nor would the seeing have been constant. Here we assume
a Gaussian resolution element with a full-width-half-maximum (FWHM)
of 6.5 km s$^{-1}$, which is a reasonable approximation for
both the HIRES and UVES data. Most \mlya\ features have
Doppler parameters $b \gtrsim 15$ km s$^{-1}$, and 
for sample inclusion we choose $b > 8$ km s$^{-1}$
(corresponding to FWHM 13.3 km s$^{-1}$
), so even 10\% uncertainties in the instrumental
FWHM do not make a significant difference.
We therefore convolve all simulated spectra with a Gaussian
kernel with FWHM=6.5 km s$^{-1}$.

\item The continuum estimates are based on large scale high
points in the observational data, and may be in error. {\sc VPFIT}
allows a linear continuum adjustment as a function of
wavelength over the fitting region, and inserts this adjustment
automatically when the overall fit accuracy comes down
to a specified level. To be consistent this was used both for
the observational data and for the simulations.
\item {\sc VPFIT} occasionally introduces very large Doppler parameter
lines which appear to be better described as long
range continuum adjustments. To remove these we omitted
features with Doppler parameters $> 100$ km s$^{-1}$.

\item The zero level may be offset by a small amount in the
observed spectra while it is accurately known in the simulated
ones. A zero level adjustment was introduced during the automatic fitting process when the normalized $\chi^2$ became $< 5$ if there where five or more contiguous pixels of the fitted profile below $5\times 10^{-3}$ of the continuum value. The details are given in sections 6.5 and 7.4 
of the {\sc VPFIT} documentation \citep{vpfit}.

\item Heavy element lines contaminate the \mlya\ forest in the
observational data, but are absent in the simulations. We
identified as many as we could from systems which showed
heavy element lines longward of the \mlya\ emission and then
chose wavelength regions within the forest to avoid the
stronger ones. Heavy element lines are usually narrow, so
the 8 km s$^{-1}$ threshold adopted for this analysis will remove
most of the ones we have not identified. Also in the cases where 
metal lines are clustered in groups, they are still 
fitted as separate narrow components for the signal-to-noise 
ratios in our data sample.

\item The simulated spectra cover a fixed small range of just
over 1000 km s$^{-1}$
(or 15.5 \AA\ at redshift z = 2.75), and all
lines in each of these spectra were fitted simultaneously. 
In the observed data, {\sc VPFIT} was applied to regions 
of varying size, depending on the
local line density and the positions of heavy element lines,
but were chosen to be between 10 and 25 \AA\ long,
with an average of $\sim 16$ \AA.

\item The flux noise in the observational data is not
constant from object to object, or even within an object
spectrum, where it depends on the signal. The continuum
level noise, $\sigma_c$, may be estimated by interpolating between
regions of the spectrum where there is little absorption, and
the zero level noise, $\sigma_0$, by doing so between saturated \mlya\
lines. Then for the simulations the noise can be set using
$\sigma = \sqrt{\sigma_0^2 + F (\sigma_c^2-\sigma_0^2)}$, 
where $F$ is the transmitted flux normalized by the continuum.
We identified 53 spectral sections characterised by
different ($\sigma_c, \sigma_0$) pairs. 
To account for this, the  simulated sight-lines 
are divided into  53 subsets
with path length proportional to the path lengths of the 53
sections. To each of them we add flux-dependent Gaussian noise
using the appropriate value of $\sigma_c$ and $\sigma_0$.

\item Since flux noise is estimated on a 
pixel-by-pixel basis, we rebin the simulated
flux in pixels of 2.5 km s$^{-1}$, which is the pixel scale used in most 
of the data sample. 
For simplicity, in the few cases where the pixel size, $\Delta v$, of the data is not
2.5 km s$^{-1}$, $\sigma_0$ and $\sigma_c$ of the corresponding 
simulated sets are rescaled by $\sqrt{\Delta v/2.5}$ before
adding noise. We have checked, for a sample of spectra, 
that this rescaling procedure produces consistent
fits with those in which the same pixel size and noise level as 
the data was used. 

\item {\sc VPFIT} adds as many components as necessary until
a satisfactory fit is obtained. However, it may fail to converge
to give an acceptable fit to either observational or
model data, sometimes e.g. if there is a saturated \mlya\ line.
Where this happens the fits from those regions are omitted.
Model spectral regions were chosen with noise 
characteristics mimicking the observational
ones with acceptable fits and, in cases where convergence
failed, different sightlines through the model were
chosen until an acceptable fit was obtained.
\end{itemize}

All fitted components are characterized by their central
redshift $z$, column density $N_{HI}$ (in cm$^{-2}$)
and Doppler parameter $b$ (in km s$^{-1}$), 
and {\sc VPFIT} provides estimates and
uncertainties for all these quantities. The observational data
yielded 2271 fitted \mlya\ components in a total path length of 
$\Delta z= 2.788$.
A potential problem with the approach taken to avoid
metal line contamination is that possible CIV and MgII doublets
may occur without counterparts redwards of the \mlya\
emission line and would not be identified inside the forest.
To assess the impact of these contaminants we have compiled
a sample of lower-redshift quasars ($2.13 < z_{\rm em} < 2.54$)
and identified CIV and MgII doublets lying in the range
4316-4802 \AA, i.e. the range covered by the \mlya\ data \citep[the
same data were recently analysed by][]{Kim2016}. For
consistency, we only consider doublets with no associated
lines from lower ions (and longer wavelength transitions),
which would have been detected in the first place by our
identification process. The Voigt-profile fit parameters from
this sample is then used to add the opacity of these contaminants
to the simulated spectra and estimate the impact on
our results. We have checked that by doing so the results
of this paper are not significantly affected, once our chosen
cuts on $b$ and $N_{HI}$ are applied (see below).

\section{Models}\label{Sec:hydro}

\begin{figure*}
\includegraphics[width=\textwidth]{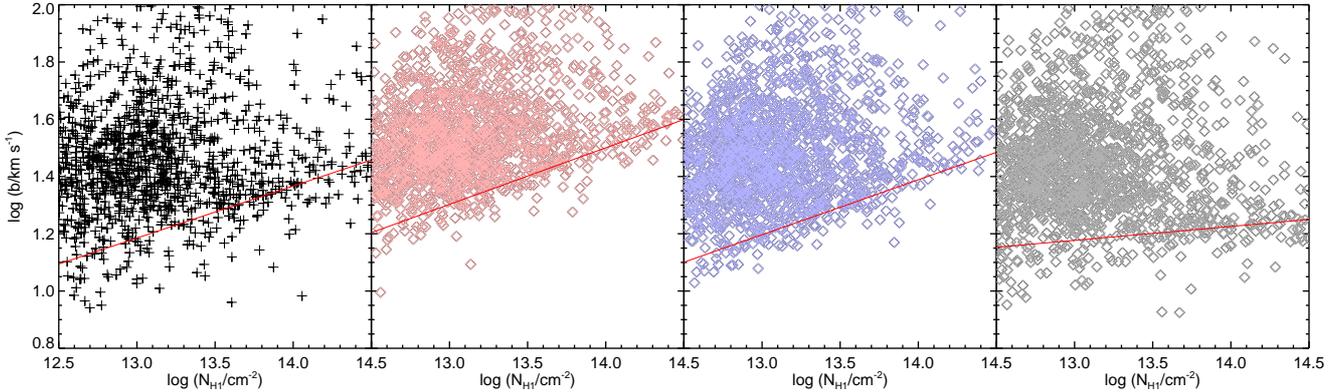}
\caption{\label{Fig:line_space} 
Distributions in the $\log N_{\rm HI}$-$b$ plane 
for the lines fitted in the quasar data sample (left) and
in three IGM models with $T_0=25000$ K, $\gamma=1.6$ (center-left), 
$T_0=15000$ K,$\gamma=1.6$ (center-right) and 
$T_0=15000$ K,$\gamma=1$ (right). All three models have 
smoothing parameter $\xi=0.8$, corresponding to a  smoothing length 
$\lambda_P=93$ kpc (comoving). The red solid lines 
represent the fitted power law to the cut-off
of the distribution, which is sensitive to the thermal 
parameters of the IGM. The behaviour of these lines in the three
models illustrates the sensitivity of the cut-off to the 
thermal parameters which has been used in the past to 
constrain the thermal parameters \protect\citep[e.g.][]{Schaye1999,Rudie2012} 
the difference in $T_0$ between the 
second and the third panels determines a clear change in the 
intercept of the fitted line. Conversely the lower slope of the 
TDR used in the right panel, which shows an isothermal model,
 makes the cutoff much flatter than in the other two cases.}
\end{figure*}

In order to predict the observed statistical properties of the 
\mlya\ forest, we used simulated spectra from the set of 
hydrodynamics simulations described in \cite[][{hereafter} B11]{BeckerBolton2011}.  The simulations were run using
the parallel Tree-smoothed particle hydrodynamics (SPH) code
GADGET-3, which is an updated version of the publicly available code
GADGET-2 \citep{Gadget2}.
The fiducial simulation volume is a 10 Mpc$/h$ periodic box
containing $2 \times 512^3$ gas and dark matter particles. This resolution
is chosen specifically to resolve the Ly$\alpha$ forest at redshift $z\sim 4-5$
\citep{Bolton2009}. The simulations were all started at 
$z = 99$, with initial conditions generated using the transfer function of \citep{Eisenstein99}. 
The cosmological parameters are $\Omega_m=0.26,\Omega_{\lambda}=0.74,\Omega_{b}h^2
=0.023,h=0.72,\sigma_8 = 0.80, n_s = 0.96,$
consistent with constraints of the cosmic microwave background from 
WMAP9 \citep{Reichardt2009,Jarosik2011}. The IGM is assumed
to be of primordial composition with a helium fraction by mass
of $Y=0.24$ \citep{Olive2004}. The gravitational softening
length was set to 1/30th of the mean linear interparticle spacing and
star formation was included using a simplified prescription which
converts all gas particles with overdensity $\Delta = \rho/\bar{\rho}>10^3$ and
temperature $T < 10^5$ K into collisionless stars. In this work we will only
use the outputs at $z=2.735$ .

The gas in the simulations is assumed to be optically thin and in
ionization equilibrium with a spatially uniform ultraviolet background
(UVB). The UVB corresponds to the galaxies and quasars emission model
of \cite{HM01} ({hereafter} HM01). Hydrogen is reionized at $z=9$ and
gas with $ \Delta \lesssim 10$ subsequently follows a tight power-law
temperature-density relation, $T=T_0 \Delta^{\gamma-1}$, where $T_0$
is the temperature of the IGM at mean density \citep{HuiGnedin97,
Valageas2002}. As in B11, the photo-heating rates from HM01 are
rescaled by different constant factors, in order to explore a variety
of thermal histories.  Here we assume the photo-heating rates
$\epsilon_i=\xi \epsilon_i^{HM01}$, where $\epsilon_i^{HM01}$ are the
HM01 photo-heating rates for species $i=$[HI,HeI,HeII] and $\xi$ is a
free parameter.  Note that, different from B11, we do not
consider models where the heating rates are density-dependent.  In
fact, we vary $\xi$ with the only purpose of varying the degree of
pressure smoothing in the IGM, while the TDR is imposed in
post-processing. 
In practice, we only use the hydrodynamics
simulation to obtain realistic density and velocity fields. 
For this
reason, we will refer to $\xi$ as the 'smoothing parameter'.  
We then impose a specific temperature-density relationship on top of the
density distribution, instead of assuming the temperature calculated
in the original hydrodynamics simulation. \review{This means that in our models 
the temperature is only a function of the density, 
strictly following eqn.~\ref{eqn:TDR}
at all densities up to $\Delta = 10$. As done in \cite{Rorai2017}, we 
set the temperature to be  constant at higher densities, i.e. $T(\Delta > 10) 
= T (\Delta = 10)$, in order to avoid unphysically high values.} Note however 
that such densities correspond to strongly saturated \mlya\ absorbers, which
are not used in our analysis (see below).  We opt for this strategy in
order to explore a wide range of parametrizations of the thermal state
of the IGM, at the price of reducing the temperature density  diagram of the gas to a
deterministic relation between $T$ and $\rho$.
In this work, we use a total of 107 models based on hydrodynamics simulations
with $\xi=0.3,0.8$ and $1.45$. The grid of parameters spans values between
0.4 and 1.9 for $\gamma$ and between $5000$ K and $35000$ K for $T_0$.

Finally we calculate the optical depth to \mlya\ photons for a set of
1024 synthetic spectra in each model, assuming that the gas is optically thin,
taking into account peculiar motions and thermal broadening. 
We scale the UV background photoionization rate in order to match 
the observed mean flux of the forest at the
central redshift of the sample\footnote{\review{In reality the 
average transmission of the \mlya\ forest evolves throughout this 
redshift bin. We have verified, however, that modeling the forest using a single value for the mean flux over this redshift range has only a small effect on the results when compared to the statistical uncertainties (see appendix~\ref{appendix}).}}
\citep[$\bar{F}_{\rm
obs}(z=2.75)=0.7371$, ][]{Becker2013}.  

We stress that in this scheme the pressure smoothing and the
temperature are set independently.  While not entirely physical, this
allows us to separate the impact on the Ly$\alpha$ forest from
instantaneous temperature, which depends mostly on the heating at the
current redshift, from pressure smoothing, which is a result of the
integrated interplay between pressure and gravity across the whole
thermal history \citep{GnedHui98}.

\subsection{Parameterization of the pressure smoothing}\label{Sec:freal}
Varying the smoothing parameter $\xi$ allows us to test the effect of
different thermal histories on the structure of the IGM density field. 
In order to characterize  it in a model-independent way, we 
adopt the definition proposed by \cite{Kulkarni2015} for the pressure
smoothing length in hydrodynamics simulations $\lambda_P$. 
This is based on the \emph{real-space} \mlya\ flux $F_{\rm real}$, 
calculated as the transmitted flux of \mlya\ photons 
in the fluctuating Gunn-Petersson approximation
\begin{equation}\label{eqn:freal}
F_{\rm real}(x)= \exp \left[-\frac{3\lambda^3_{\alpha}\Lambda_{\alpha}}{8\pi H(z)} n_{HI} \right]
\end{equation}
where $\lambda_{\alpha}= 1216$\AA\ is the rest-frame \mlya\ wavelength,
$\Lambda_{\alpha}$ is the Einstein $A$ coefficient of the transition, $H(z)$
is the Hubble parameter and $n_{HI}(x)$ is the neutral hydrogen number
density at the point $x$. Critically, $F_{\rm real}$ does not include the effects of thermal broadening or peculiar velocities on the transmitted flux.
In the optically-thin regime, where $n_{HI} \propto \rho^2$
this is a non-linear transformation of the  density field
that suppresses high densities, but preserves isotropy.
In \cite{Kulkarni2015} it is shown that the $F_{\rm real}$ power spectrum 
is sensitive to the thermal history of the low-density IGM, and 
is well approximated by the function 
\begin{equation}\label{equation:freal_fit}
\Delta_F(k)= A k^n \exp[-(k\lambda_{P})^2] ,
\end{equation}
where $k$ is the Fourier wavenumber and 
$A$, $n$ and $\lambda_{P}$ are the free parameters.
We calculate the $F_{\rm real}$ field for our hydrodynamics
simulations, using the $n_{HI}$ output at $z=2.735$. Consistently to what we 
have done for the post-processed models described in the previous section,
we rescale $n_{HI}$ so that the mean flux in a set of synthetic \mlya\ spectra
extracted from the simulations matches the observed values at the correspondent redshift.
We obtain
$\lambda_P$ from the fit of the power spectrum. This gives
$\lambda_P=69,93$ and 108 kpc for $\xi=0.3,0.8$ and 1.45,
respectively. \review{Here and elsewhere in the paper we always express
the smoothing length in comoving kpc.}

\section{Methodology}
\label{Sec:method}

\begin{figure*}
\includegraphics[width=\textwidth]{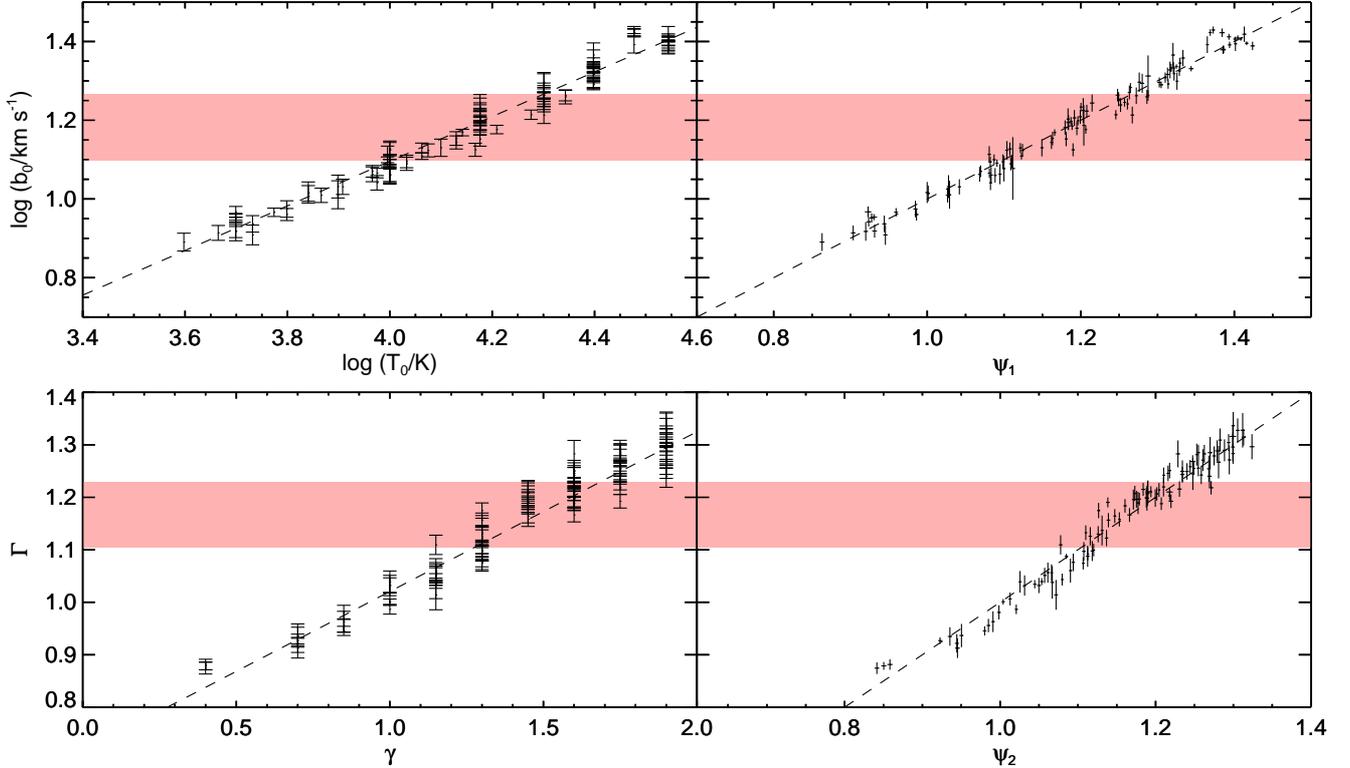}
\caption{\label{Fig:calibration_original}
Calibrated relations between the parameters $b_0$ and  $\Gamma$ describing the normalisation and slope of the  
lower cutoff in the $b-N$ distribution  and the thermal parameters $T_0$ and $\gamma$ describing the TDR. 
\emph{Left column:} cutoff normalization 
$\log b_0$ at $\log N_{\rm HI,0}/{\rm cm}^{-2}=13$ 
as a function of  $\log T_0$ (upper panel) and slope of the fitted 
 $\Gamma$ as a function of the index $\gamma$  of the TDR (lower panel). 
Each point represents a different model in the grid and errors are 
estimated by bootstrapping the simulated line samples. The red bands delimits 
the values of $\log b_0$ and $\Gamma$ measured from the data within 
the 1-$\sigma$ error. 
The calibrated function $\log b_0(\log T_0)$ is obtained via a 
linear fit of these points (dashed lines).
The two relations shown in this plot are reasonably tight, but the level
of scatter appears higher than what the error bars warrant, suggesting
that dependencies on other parameters are not negligible.  This can
have an effect on the inference of the thermal parameters.
\emph{Right column:} same as the left panel, but using the generalized
combination of thermal parameters described in \S~\ref{sec:optimal_comb}.
Each point represents the mean of the MCMC posterior distribution for 
the combinations used to describe $\log b_0$ (upper panel) and 
$\Gamma$ (lower panel). The (small) horizontal error bars represent 
the propagated uncertainties on the combination coefficients as obtained from the 
posterior distribution. The dashed lines represent the identity.
Adopting the generalized scheme substantially improve the fits: 
the chi-square for the four plots, divided by the number of points, 
is 3.49 (top left), 2.70 (top right), 4.16 (bottom left) and 2.56 (bottom
right)}
\end{figure*}

Simulated spectra for various IGM models were produced
to be compared with the data, and we derived Doppler parameter
 $b$ and column densities $N_{HI}$ for the \mlya\ lines using {\sc VPFIT}
in exactly the same way as for the telescope data (see \S~\ref{Sec:data}
and \S~\ref{Sec:hydro}). The distribution of lines in the 
$b-N_{HI}$ plane forms the basis for our statistical analysis.
For both the data and the simulations there may be 
velocity structure which is not well represented by Voigt profiles. 
In the fitting process this can result in a number of
non-physical components, which can be quite close together,
or in blended, broad but weak lines. A
feature of these is that the error estimates tend to be large as
a consequence of the presence of neighbouring systems.
An example can be seen in Fig.~\ref{Fig:vpfit},
showing the results of applying the algorithm to a simulated 
spectrum. To remove many
of these, and to prevent the $b-N_{HI}$ distributions being 
dominated by noise, we require the relative error estimate
in the Doppler parameter to be smaller than 50 \%, and
the error on $\log N_{HI}$ smaller than 0.3.
Additionally, for $\log N_{HI} > 14.5$ the \mlya\ line is usually
saturated to a level where the column densities derived
from fitting only the \mlya\ transition are unreliable. So we 
exclude systems with higher column densities from the analysis. 
Finally, for the highest $b$-values in the range the line
detection limit is $\log N_{HI} \sim 12.5$, so we adopt this as a lower
limit for the comparison samples.
To summarize, our statistical analysis is based on the absorption
components with $8<b<100$ km s$^{-1}$, 
$12.5 < \log (N_{HI}/ {\rm cm}^{-2}) < 14.5$,
 Doppler parameter relative error smaller than 50\% and log column
density error < 0.3. With these
restrictions the observational data sample consists of 1625
points in the $b-N_{HI}$ space.
The line distribution for the  data and for three different models
is shown in Fig.~\ref{Fig:line_space}.

\subsection{The lower cut-off in the $b-N$ distribution}
As noted previously 
\citep{Schaye1999,Theuns00,McDonald2001}  the $b-N$ distribution 
shows a pronounced cut-off at
low values of $b$. This is usually interpreted as a signature of thermal
broadening setting a lower limit to the absorption line widths. 
The position of this cutoff is dependent on the 
column density, suggesting that the  temperature systematically varies with the 
density of the gas. This  motivated several studies to use  the slope and normalisation  of 
the lower of the $b-N$ distribution as a direct probe of the TDR of the IGM. 
Pressure smoothing also broadens the lines by increasing the 
physical size of absorbers, which increases their velocity width 
due to the Hubble flow. 
This effect can be taken into account by means of theoretical/analytical 
arguments or hydrodynamics simulations \citep{Schaye1999,Theuns00,Garzilli2015}.

The lower cutoff in the $b-N$ distribution 
is generally assumed to be a power-law relation
between line width and column density,
\begin{equation} 
b=b_0 \left(\frac{N_{\rm HI}}{N_{\rm HI,0}}\right)^{\Gamma -1},
\label{eqn:cutoff}
\end{equation} 
where $b_0$ and $\Gamma$ are the parameters connected with the TDR and
$N_{\rm HI,0}$ is a reference value which is 
often chosen as the column density corresponding 
to gas at the mean density $\Delta=1$ 
\citep[see][for a discussion]{Bolton14}. In Fig.~\ref{Fig:line_space} we show the fitted cutoff for our observed absorption line
sample and samples of simulated absorption lines for three different thermal models. In our analysis  we arbitrarily  fix 
$N_{\rm HI,0}=10^{13}$ cm$^{-2}$, but we will relax the assumptions on 
the functional dependencies between $b_0,\Gamma$ and the thermal parameters.
The standard algorithm used to fit this cutoff, which we 
also adopt, was first introduced by
\cite{Schaye1999} and is based on a recursive rejection process:
the expression in eqn.~\ref{eqn:cutoff} is fitted to the line distribution
to obtain $b_{\rm fit}(N_{HI})$. We then calculate the 
mean absolute deviation $|\sigma_b|$ of the points from the fit in the 
$b$ dimension, and discard all lines whose value of $b$ is \emph{greater}
than the fitted relation by more than $|\sigma_b|$. The fit and the 
rejection of upper outliers are iteratively repeated until 
convergence, i.e. when all points lies below the upper mean deviation
and the only outliers are below the fit. At that stage, the lines with 
$b < b_{\rm fit}(N_{HI}) - |\sigma_b|$ are discarded and the 
fit is repeated one last time to define the final
values of $b_0$ and $\Gamma$.
The errors associated with $\log b_0$ (it is convenient to operate
in logarithmic space) and $\Gamma$ are estimated 
via a bootstrap technique applied to the line sample.  In a first 
approximation we follow the standard practice of 
considering $\log b_0$ and $\Gamma$ as statistically independent 
and treat them separately. Correlations between these parameters are addressed in the next section.

\review{One may notice from Fig.~\ref{Fig:line_space} that the 
lines fitted in the data sample (left panel) present more 
outliers below the cut-off compared to
the models (right three plots). To understand what could 
generate this difference, we visually inspected all individual
absorbers in the data 
with width lower than the cut-off by more than $\log b=0.1$ .
Among 29 outliers, we found 7 that are compatible with being
unidentified metal lines and other 7 which are \mlya\ absorbers
associated with metals at the same redshift. The latters are 
metal-enriched systems whose temperature might be affected
by metal cooling, which is not accounted for in our models. 
We have tested the effect of removing these 14 
lines and to re-apply the 
cut-off fitting procedure. This did not significantly change the
estimated cut-off parameters, but it reduced the uncertainty on 
$\log b_0$ by about 20\%. For this reason we consider keeping all the
outliers to be the most conservative choice.}

In the published literature, the relations between the narrow line 
cut-off parameters and the thermal parameters 
($T_0,\gamma$ and in this work also $\lambda_P$),
were based on theoretical arguments \review{set out by  \cite[][, see also \cite{Rudie2012}]{Schaye1999} }
or calibrated with hydrodynamics simulations \citep{Bolton14}. 
Here we will first make the ansatz that such relations can 
be written in the form   
\begin{eqnarray}\label{eqn:calibration}
\log T_0=A + B \log b_0 \\
\label{eqn:calibration_2}
\gamma -1 = C + D (\Gamma -1) 
\end{eqnarray} 
where $A,B,C,D$ are determined using the full set of 
models described in \S~\ref{Sec:hydro} (see
Fig.~\ref{Fig:calibration_original}). Analogous to 
previous works, we for now assume no dependency on the 
pressure smoothing length $\lambda_P$, which for standard hydrodynamics simulations 
implicitly assumes 
that the thermal history is correct and hence  
consistently taken into account. Note that these relations are slightly
different than what is usually assumed in other works.
We will later check/verify the validity of this ansatz \emph{a posteriori} 
using the model grid (see Fig.~\ref{Fig:calibration_original}). 

In the next section we discuss how 
we abandon these assumptions in order to include a more 
general relation between the parameters describing the lower cutoff 
of the $b-N$ distribution  and parameters describing the thermal state and history of the IGM, 
including the smoothing length $\lambda_P$.

\subsection{Generalising the  calibration of the cut-off parameters}
\label{sec:optimal_comb}

\begin{figure*}
\includegraphics[width=\textwidth]{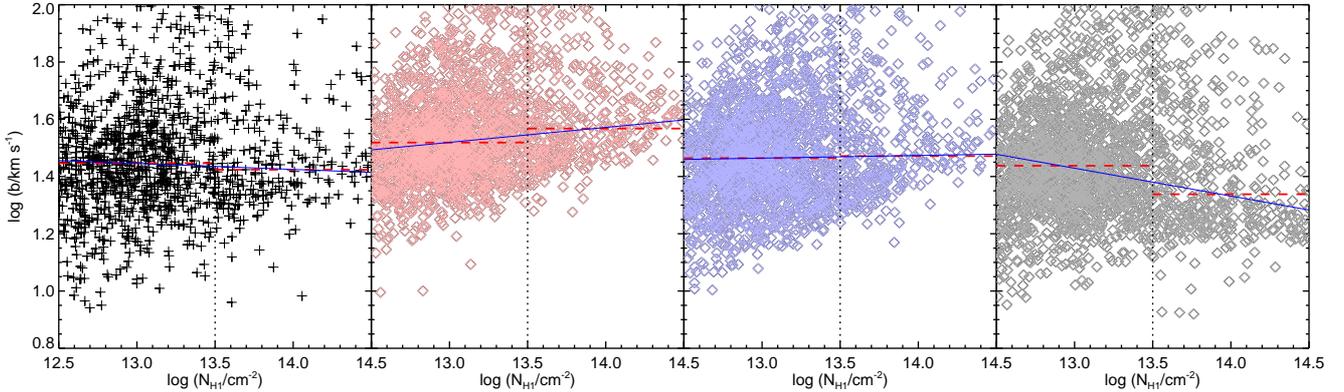}
\caption{\label{Fig:median_statistic}
Representation of the differential median statistic. The points in the
four panels show the  $b-N$ distribution of the data
and the simulation for the same three thermal models
as in  Fig.~\ref{Fig:line_space}. The vertical dotted lines separate the two
$N_{HI}$ ranges for which the two medians $m_1$ and $m_2$ (shown in 
dashed red horizontal lines) are calculated. The blue lines describe 
the equation $b=m_1'-\alpha(\log N_{HI} - 13)$ chosen so  that they
horizontally divide both column-density ranges into two parts with 
 an equal numbers of points (see text for details). $m_1$ and $\alpha$
 are the two parameters used in the analysis. The plot illustrates their
sensitivity to the thermal parameters $\gamma$ and $T_0$. }
\end{figure*}

The relations described by \review{eqns.~\ref{eqn:calibration} and 
\ref{eqn:calibration_2} are based} 
on the heuristic argument that the renormalization of the TDR 
changes the line widths at all column densities by a similar proportion,
at least for lines that fall near the cutoff. Although Fig.~\ref{Fig:calibration_original}
suggests that such an approximation is reasonable, we would like to 
consider the possibility of more general dependencies in order to find
a scheme to quantify the effect of the pressure smoothing
on the line distribution.

We approach this problem by starting with  the assumption that a generic observable
$\phi$ can be approximated by a linear combination of the
logarithm of the relevant parameters. More precisely, we define
the combination $\psi$ as
\begin{equation}  \label{eqn:linear_comb}
\psi = a + b \log T_0 + c (\gamma - 1) + d \log\lambda_P
\end{equation} 
where $a,b,c,d$ are free parameters. The choice of using
logarithmic  quantities (except for $\gamma$, which is an index) 
is equivalent to assuming  a power-law relation
with $T_0$ and $\lambda_P$. Note that changing the units in which 
$T_0$ and $\lambda_P$ are expressed would only affect the offset $a$.
In general, $\phi$ could be
a more general function of the parameters than a
linear relation, so the validity of this approximation will need to be
verified \emph{a posteriori}. 
If $\psi$ is not a good description of the observable, 
or if there are statistical uncertainties in the models (e.g. due to finite 
box size), it is natural to expect some scatter. This
can be accounted for by introducing some flexibility in choosing
$\{a,b,c,d\}$. 

A convenient way of doing so is to implement a 
likelihood probability for fitting $\phi$ with $\psi$ and to include 
it in a Markov-Chain Monte Carlo (MCMC) algorithm in the 
space of coefficients. This can be combined with a likelihood
of the measured observable $\phi_d$ when compared to the
value of $\psi$ for a given choice of the coefficients and
of the thermal parameters. This allows us to draw quantitative
inferences on the calibration coefficients $\{a,b,c,d\}$ 
and on the thermal parameters $\{T_0,\gamma,\lambda_P\}$ at the same 
time.

In our analysis the total likelihood is therefore composed of  two parts. The first
part quantifies how well the parameter combination $\psi$ defined by 
$a,b,c,d$ fits the points in the training grid of models.
Given the observable  $\phi$, this can  be written as
\begin{equation}\label{eqn:like1}
\log L_1=-\sum_i \frac{[ a + b \log T_{0,i} + c 
(\gamma_i - 1) + d\log\lambda_{P,i} - \phi_i ]^2}{2 \sigma^2_{\phi,i}}
\end{equation}
where the sum is performed over all the models in the grid and
$\sigma_{\phi,i}^2$ is the uncertainty on  $\phi_i$, both relative 
to the $i$-th model. An MCMC run which employs 
$L_1$ as likelihood  suffices to estimate the posterior
distribution for the calibration coefficients $a,b,c,d$, i.e.
the range of parameter combinations which can be used to 
relate the observable $\phi$ to the IGM parameters via
eqn.~\ref{eqn:linear_comb}. In principle, this distribution could be used
as a prior for a second MCMC run which includes  the thermal parameters. 
However, we find it more practical to  combine $\log L_1$ with  
 a likelihood where the same values of $a,b,c,d$ are employed to make a prediction for 
$\phi(T_0,\gamma,\lambda_P)$ and test it against the data.
Assuming  the error is Gaussian, this can be  simply expressed as 
\begin{equation}\label{eqn:like2}
\log L_2= - \frac{[a + b \log T_0 + c (\gamma - 1) +
d\log\lambda_P - \phi_d ]^2}{2\sigma^2_{\phi,d}},
\end{equation}
where $\phi_d$ and $\sigma_{\phi,d}$ are estimated from the 
data. The two parts of the likelihood can then be summed to 
obtain the total likelihood which we use to run a MCMC
in the 7-dimensional space of the calibration and thermal
parameters \{$a,b,c,d,T_0,\gamma,\lambda_P$\}.  

This scheme is computationally inexpensive and 
achieves several goals:
\begin{itemize}
\item It generalizes the simple
power-law calibration assumed in eqn.\ref{eqn:calibration}.

\item It finds the parameter combinations to which 
the observable is sensitive, providing a way to 
quantitatively express its degeneracy direction.

\item The uncertainty of this  calibration can be determined based on
the scatter of the simulated points and their uncertainties.

\item It returns the constraints on the thermal parameters, 
marginalizing over the uncertainty on the coefficients of 
eqn.~\ref{eqn:linear_comb}.	
\end{itemize}
The  main drawback is that the validity of  
eqn.~\ref{eqn:linear_comb} as an approximation of the observable $\phi$ 
must be verified 
\emph{a posteriori}, using the estimated coefficients from 
the MCMC run.

In the case where the observable $\phi$ is not a scalar but a vector,
it is possible to generalize the likelihood expressions \ref{eqn:like1}
and \ref{eqn:like2} to multivariate Gaussian likelihoods. 
This will increase the dimensionality of the calculation, because 
there must be four coefficients $\{a,b,c,d\}$ for each
observable component. It will also be necessary to 
calculate the full covariance matrix of the different components
of $\phi$.

In this work, we have applied this formalism to the 
observables $(\log b_0, \Gamma)$  that describe the lower cut-off
of the $b-N$ distribution  and to the differential median parameters $(m_1,\alpha)$ 
that we will define in the  next section. In both cases this involved an 11-parameter 
MCMC analysis and  adopting a bivariate Gaussian likelihood, both for 
eqn.~\ref{eqn:like1} and \ref{eqn:like2}. The correlation between
$\log b_0$ and $\Gamma$ has been calculated by bootstrapping the 
sample of absorption lines, as it was done for the standard deviations.

\subsection{Differential median of the line width distribution}
\label{Sec:median_method}

Although the cutoff is 
certainly the most  prominent feature of the distribution,
it is reasonable to expect that the thermal properties of the 
IGM do not only affect the narrowest lines in the \mlya\ forest. 
Taking advantage of the 
additional information contained by the bulk of the 
line population is an obvious way of trying to refine 
the constraints achievable with a Voigt-profile 
decomposition approach. 
Some of our preliminary efforts to do so encountered systematic issues related to very broad lines ($b > 60$ km s$^{-1}$).
Our  models were not able to reproduce the observed line distribution in that range, 
and, more problematically, inferences on the thermal parameters 
from various fitting schemes were strongly dependent 
on the way the upper $b-$range was chosen or on
the statistic employed to characterize the line distribution. 
We concluded that the broad  lines are in most cases probably an 
artefact of the fitting procedure and do not represent physical properties 
of the IGM  (see for example Fig.~\ref{Fig:vpfit}) and are therefore  more subject to 
systematic errors, in particular due to the uncertainty of  the
continuum placement. 

For this reason, we turned our attention to a statistical estimator which is not
sensitive to the distribution of lines at extreme values
of $b$. A natural choice is the median. In order to 
capture the column-density dependency of the Doppler-parameter 
distribution, we adopt the following approach.
We define $m_1$ as the median of the line width distribution for
$10^{12.5}<N_{HI} < 10^{13.5}$ cm$^{-2}$ and $m_2$ as 
the median for 
$10^{13.5}< N_{HI} < 10^{14.5}$ cm$^{-2}$. We then apply a linear transformation
to $\log b$ parametrized as 
\begin{equation}
\log b' = \log b + \alpha \log\left[\frac{N_{HI}}{10^{13} {\rm cm}^{-2}}
\right],
\end{equation}
where  the transformation coefficient $\alpha$ is such that the 
medians $m_1'$ and $m_2'$ of the new quantity $\log b'$ are 
identical (for the same 
$N_{HI}$ domains defined above). It is straight forward to determine 
$\alpha$ iteratively. This calculation is different from simply
considering the differences of the median $m_2-m_1$, because 
$\alpha$ depends on the positions of all individual lines in the plane
and  not just on the median of $\log b$ in the two parts. However, 
for simplicity we will
refer to this method as  the 'differential median' method.
We choose $m_1$ (the median calculated \emph{before} the transformation)
and $\alpha$ as our final observables to estimate the thermal parameters,
to which we apply the  analysis technique illustrated in the previous section. 
Analogously to the cutoff method, we 
calculate uncertainties and covariance for $m_1$ and $\alpha$ 
by  bootstrapping the line sample. 
An illustration of our differential median  statistic is 
shown  in Fig.~\ref{Fig:median_statistic}, both for the data 
and for the same three thermal models shown in Fig~\ref{Fig:line_space}. \review{Its sensitivity to the parameters
of the TDR is shown by the different slopes and normalizations
of the lines for the different thermal models, which qualitatively 
have the same parameter dependencies as the cut-off.}

\section{Results}
\label{Sec:results}

\begin{figure*}
\includegraphics[width=\textwidth]{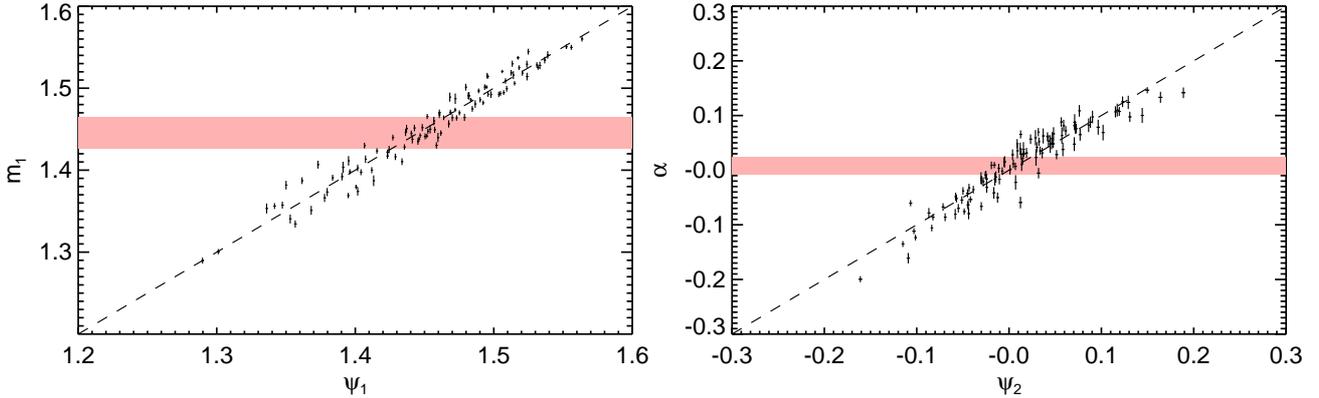}       
     \caption{\label{Fig:median_measurement} 
     Same as Fig.~\ref{Fig:calibration_original} (right column), but for 
     the differential median method, showing the relation 
     between the low-column-density median
	 $m_1$ and the differential coefficients $\alpha$ with the 
	 respective average parameter combinations obtained by the MCMC
	 analysis, $\psi_1$ and $\psi_2$. 	 
     }     
\end{figure*}

In the left-hand panels of Fig.~\ref{Fig:calibration_original} we show the values of 
$\log b_0$ in the simulations as a function of the 
temperature at mean density and $\Gamma$ as a function 
of the slope of the TDR $\gamma$. When fitting the linear relations
in eqn.~\ref{eqn:calibration}, we obtain $A=2.06, B=1.76$; 
$C=-0.07, D=3.28$ (corresponding to the black dashed lines).
The cutoff-fitting algorithm applied to the data returns 
$\log b_0= 1.186 \pm 0.084 $ and $\Gamma=0.180 \pm 0.062 $, 
marked as the  red shaded region in the plot.
By applying the above coefficients
to convert these numbers to a measurement of the TDR parameters,
we get $T_0/ 10^3\, {\rm K}=14.3 \pm 5.0 $ and $\gamma=1.52
\pm 0.22$. The reported uncertainties take into account the propagated
(small) errors of the cut-off calibration coefficients.

\begin{figure*}
\includegraphics[width=\textwidth]{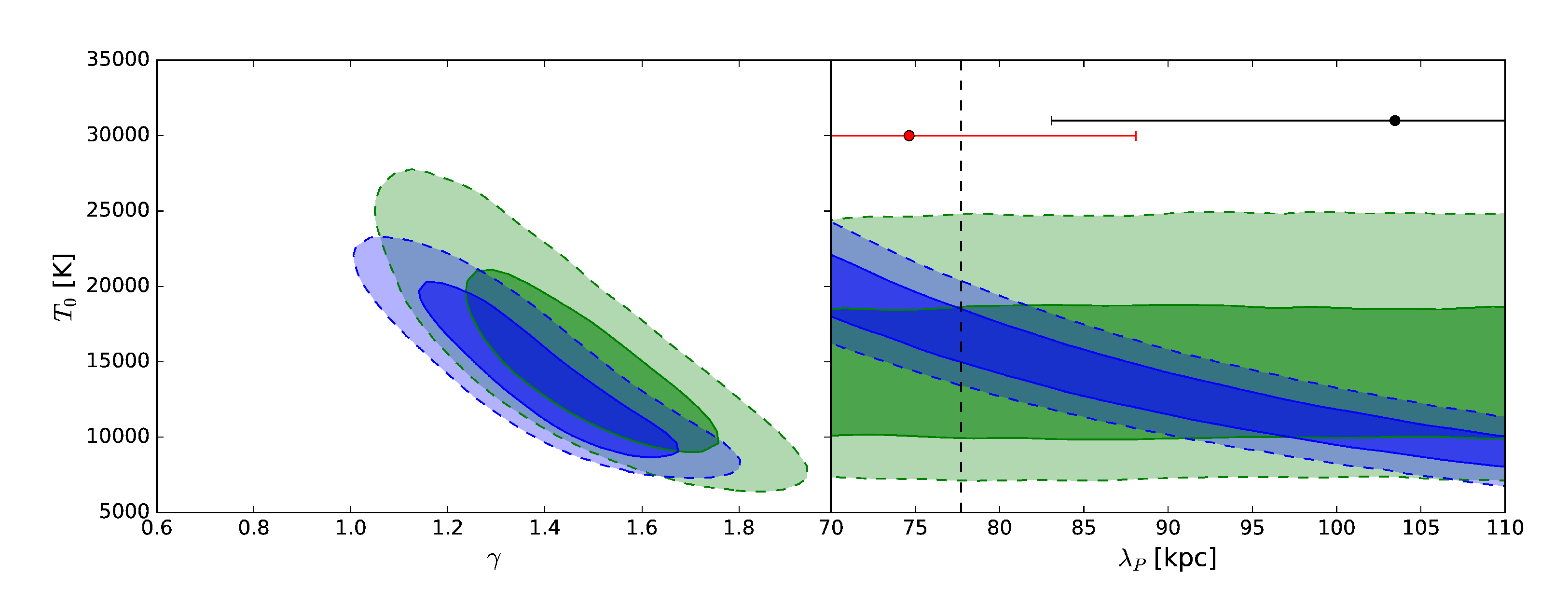}
\caption{\label{Fig:contours}Constraints on the thermal
     parameters from the two analyses presented in this work.  In green we
     show the 1-$\sigma$ (dark) and 2-$\sigma$ (light) confidence levels
	 in the $T_0-\gamma$ and $T_0-\lambda_P$ planes, derived from the cutoff
	 method. The analysis takes into 
	 account the statistical correlation between the uncertainties of 
	 the two observables, $\log b_0$ and $\Gamma$. Mainly due to this correlation, there is a strong
	 degeneracy  between the estimated values of $T_0$ and $\gamma$, although the 
	 marginalized uncertainty of the two parameters of the TDR is 
	 comparable to the one achieved in the standard method. The 
	 right panel shows that the temperature  $T_0$ inferred from the lower cut-off 
	 the $b-N$ distribution is not sensitive to the smoothing scale $\lambda_P$ 
	 of the numerical simulations used for calibration. The blue contours are 
	 the 1-$\sigma$ (dark) and 2-$\sigma$ (light) confidence levels
	 for the differential median method. Similar to the   
	 analysis based on the lower cut-off of  the $b-N$ distribution, there is a 
	 strong degeneracy between the inferred $T_0$ and $\gamma$; however, for this 
	 method the
	 temperature is also strongly degenerate with the smoothing length
	 $\lambda_P$. The range for $\lambda_P$ shown in this plot is 
	 consistent with recent measurements from \protect\cite{Rorai2017b}
	 at $z\in 2.2-2.7$ (shown as a red circle with errorbars) and $z\in 2.7-3.3$
	 (black, the vertical position of these two points is arbitrary). The vertical
	 dashed line shows the value of $\lambda_P$ measured in the non-equilibrium
	 hydrodynamics model by \protect\cite{Puchwein2015} at $z=2.79$. 
	 Stronger constraints on the pressure
	 smoothing will eventually help break the degeneracy shown
	 in this plot and significantly improve the precision achievable with this 
	 technique. }

\end{figure*}

The generalization of the relation between the thermal parameters $T_0$ and $\gamma$ and the 
parameters characterising  the lower cut-off  of the $b-N$ distribution 
described in \S~\ref{sec:optimal_comb} is shown in the right-hand panels 
of Fig.~\ref{Fig:calibration_original}. These plots are analogous to
the left-hand panels except that the $x$-axes are
 combinations of the thermal parameters as 
defined in eqn.~\ref{eqn:linear_comb}.
The coefficients for eqn.~\ref{eqn:linear_comb} 
are picked from the mean values of the MCMC 
posterior distribution,  applying the method described in 
\S~\ref{sec:optimal_comb} to a joint analysis of $\Gamma$ and $\log b_0$. 
More precisely, the average combinations are
$\psi_1=-1.03+0.55 \log T_0 - 0.02 (\gamma -1) - 0.02 \log \lambda_P$
and
$\psi_2=0.10-0.06 \log T_0 +0.27 (\gamma -1) + 0.09 \log \lambda_P$. 
These  relations do not represent a new 'calibration' in that
the coefficients are free to vary consistently with the uncertainties
on the values extracted from the simulations. However, showing that 
the relations between the observables and the typical combinations 
from the MCMC fall close to the  identity relation (dashed line) is necessary to
validate our approach. 
The constraints on the thermal parameters derived from the same MCMC
analysis  are shown in the green contour plots in
Fig.~\ref{Fig:contours}. A degeneracy between the inferred values of
$T_0$ and $\gamma$ is evident, which is a consequence of taking 
the statistical correlation between the measured $\Gamma$ and 
$\log b_0$ into account. The right-most panel demonstrates that the lower cut-off in the simulated
$b-N$ distribution is  not sensitive to the smoothing length $\lambda_P$. 
When marginalized over all parameters of the MCMC analysis, including the coefficients
for $\psi_1$ and $\psi_2$, the  values inferred for the TDR parameters are
$T_0/ 10^3\, {\rm K} =15.6 \pm 4.4$ and $\gamma=1.45 \pm 0.17$ (quoted as mean and 
standard deviation from the posterior distribution). The uncertainties are 
smaller than those inferred from the lower cut-off, and 
the two measurements  are in good agreement.

The results obtained from the differential median technique are presented
in Fig.~\ref{Fig:median_measurement}. The two panels
show that the quantities $m_1$ and $\alpha$ are reasonably approximated 
by a combination of the form of eqn.~\ref{eqn:linear_comb}, although 
the scatter is 
not negligible. As in Fig.~\ref{Fig:calibration_original},
the red bands trace the 1-$\sigma$
limits for the measured parameters: $m_1=1.446 \pm 0.020$ (left)
and $\alpha=0.007 \pm 0.016$. The relations are well described by  
$\psi_1=-0.433+0.256 \log T_0 +0.038  (\gamma -1) + 0.409 \log \lambda_P$ and 
$\psi_2=0.285 -0.141 \log T_0 - 0.194 (\gamma -1) + 0.200 \log \lambda_P$.
The posterior distribution of the thermal parameters, shown as blue contours in 
Fig.~\ref{Fig:contours}, 
reveals degeneracies between the inferred values of all three 
considered parameters. In 
particular, and in contrast to the cutoff method, there is a significant 
sensitivity to the pressure smoothing length $\lambda_P$. This confirms that the 
the overall $b-N$ distribution contains  information about the spatial smoothing of
the IGM gas, as already noted by \cite{Garzilli2015}.
The marginalized uncertainties for the TDR parameters are in this case 
$T_0/ 10^3\, {\rm K}=14.6\pm 3.7$ K and $\gamma=1.37 \pm 0.17$, in good agreement with
those inferred from the lower  cut-off of the $b-N$ distribution. The fact that using the full  distribution does 
not substantially improve the accuracy of the constraints, compared to using
just the lower cut-off, is due to  the degeneracy with $\lambda_P$, which was negligible for the lower cut-off method. On the other hand, the emergence of such 
 degeneracy implies that the precision can be 
substantially improved by combining these results with independent 
measurement of the smoothing length.
At the moment, the only constraints available on $\lambda_P$ are those 
obtained by the analysis of correlated \mlya\ absorption in close quasar pairs
\citep{Rorai13,Rorai2017b}. The horizontal bars in Fig.~\ref{Fig:contours} 
represent the result from \cite{Rorai2017b} in the redshift bins $2.2-2.7$ (red)
and $2.7-3.3$ (black), both overlapping with the redshift range analysed in the present
work. The nominal value of the $\lambda_P$ measurement has been decreased by
10\%, in order to match the definition of pressure smoothing as given by the 
$F_{\rm real}$ cutoff \cite[see \S~\ref{Sec:freal} and Fig. S11 in][]{Rorai2017b}. 
As it can be seen, the whole range for $\lambda_P$ considered in this paper
is consistent with either of the two data points. More precise 
measurements of the smoothing, or an analysis conducted on the same redshift limits
used in this work,
are required in order to break the degeneracy and improve the accuracy on $T_0$
and $\gamma$ achievable with the differential median technique.

\begin{figure}
\includegraphics[width=\columnwidth]{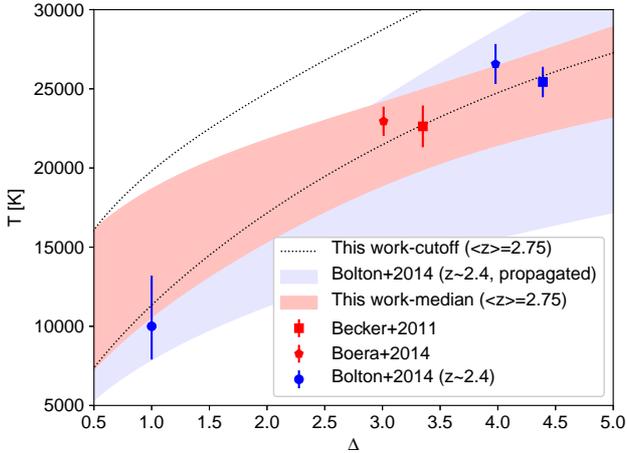}
\caption{\label{Fig:temperature_density}
Constraints on the TDR in the temperature-density plane. The black dashed lines 
show the 16th and 84th percentile of the temperature
posterior distribution as a function of density,  obtained
from our analysis of the lower cut-off of the  $b-N$ distribution. 
Analogous limits for the differential median method are 
shown by the red shaded area. The two measurements
are in good agreement with each other. The red square and pentagon
report the measurements of $T(\Delta)$ at the same redshift from
BB11 and \protect\cite{Boera2014}, respectively.
For comparison, we also report the results for $T_0$ at $z=2.4$ 
from \protect\cite{Bolton14} (blue circle) and the propagated 1-$\sigma$
limits at all densities considering the measurement of $\gamma$
from the same work (light blue shaded area). The blue square and pentagons are the 
values of $T(\Delta)$ obtained in BB11 and
\protect\cite{Boera2014} at this redshift.}
\end{figure}

An alternative way of visualizing our results is 
to look at the constraints in the temperature-density plane. This 
does not require us to reduce the posterior distribution to two parameters with relative
uncertainties. It also allows a straightforward comparison with 
the measurement of the temperature at overdensity $\bar{\Delta}$ 
obtained with the 'curvature' statistic in BB11 and \cite{Boera2014}, 
without
requiring error propagation by  which some information might be lost. 
The way we do this is the following: given the posterior distribution for
$T_0$ and $\gamma$ obtained from one of our MCMCs, we calculate the
marginalized distribution of $T_0 \Delta^{\gamma -1}$ for the 
range of overdensities $\Delta$ we are interested in. We then 
plot the 16th and the 84th percentile as a function of $\Delta$
as the 1-$\sigma$ limits in the temperature-density 
plane. The results for the various techniques described in this paper
are  shown in Fig.~\ref{Fig:temperature_density}. The black dotted
lines show  the 16th and 84th percentiles of  the temperature 
distributions obtained from the lower cutoff MCMC analysis, while the red
shaded region shows the same for the differential median method. 
 Note that the curvature results do not include uncertainties related to pressure smoothing. There 
is good  agreement between the two measurements, and there is also good 
consistency with the results from BB11 and \cite{Boera2014} at the same redshift
(red square and pentagon, respectively), although our dataset partially overlap with 
the sample used in \cite{Boera2014}.
Note that the uncertainties of the temperature 
measured with the median technique is lower at mild overdensities ($\Delta \sim 2.5-3.5$) than at the mean density. 
This is a consequence of the particular degeneracy direction in the 
$T_0-\gamma$ plane of Fig.~\ref{Fig:median_measurement}. 

For reference
we also show an analogous comparison at $z=2.4 $ between the results of 
BB11 (blue square), \cite{Boera2014}(blue pentagon) and $b$-$N_{HI}$ cutoff 
results from \cite{Bolton14}(blue circle). These limits assume that $T_0$ and $\gamma$ are uncorrelated, which is likely 
 incorrect given the results of our analysis at slightly higher redshifts.
The light blue shaded area in the background represent the propagated 
1-$\sigma$ limits on $T(\Delta)$ assuming the measured value and uncertainties
of $T_0$ and $\gamma$ from \cite{Bolton14}.

\section{Discussion}
\label{Sec:discussion}

\begin{figure}
\includegraphics[width=\columnwidth]{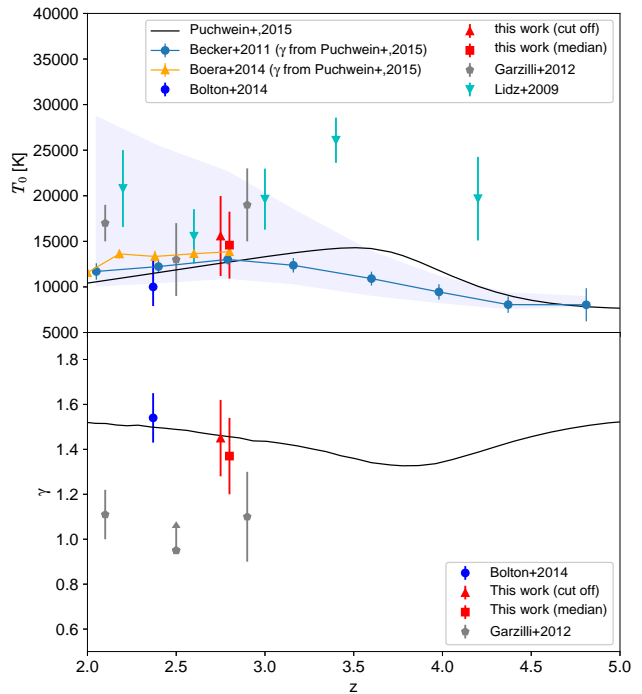}
\caption{\label{fig:main_result}
A summary of recent constraints on the TDR parameters from the 
literature, as a function of redshift. 
Our constraints on $T_0$ and $\gamma$ derived from the 
lower  cut-off of the  $b-N$ distribution (red triangles) and the differential median (red
squares) methods are compared to the hydrodynamics model of 
\protect\cite[black line]{Puchwein2015} and the extrapolated values of 
$T_0$ from the curvature measurement of BB11 (dark blue connected circles).
The latter values are obtained from the temperatures at a redshift-dependent
overdensity $\bar{\Delta}(z)$  by assuming the value of $\gamma$ from 
the \protect\cite{Puchwein2015} model (black line in the lower panel). The light-blue shaded
area represents the range of values extrapolated from BB11 assuming a 
flat prior on $\gamma$ in the interval $1 - 1.6$. We also plot the 
constraints from \protect\cite[blue circles]{Bolton14}  at somewhat lower redshift. 
The orange triangles are the values of $T_0$ 
extrapolated from a recalibrated version of the curvature measurement 
from \protect\cite[and Boera, private communication]{Boera2014}, obtained
in the same way as for BB11. The cyan triangles show the constraints 
on $T_0$ from \protect\cite{Lidz09} and the grey pentagons those from
\protect\cite{Garzilli2012}, both obtained with a wavelet analysis (combined with
the flux PDF in the latter). The grey pentagons in the lower panel 
are also from the PDF-Wavelet combined analysis of
\protect\cite{Garzilli2012}. The tension between the points at $z=2.9$ with
our measurements  is discussed in the text.}
\end{figure}

A more comprehensive comparison of the main results of this work with 
recent constraints on the thermal state of the IGM from the literature is presented 
in Fig.~\ref{fig:main_result},
where we show the evolution of $T_0$ (upper panel) and $\gamma$ (lower panel)
as a function of redshift. The red triangles correspond to the  constraints
from our fit to the lower  cut-off  of the  $b-N$ distribution (\S~\ref{sec:optimal_comb}), while the red squares are those 
obtained from the differential median method (\S~\ref{Sec:median_method}).
The black solid lines are the predictions of a recent hydrodynamics
simulation from \cite{Puchwein2015} where the UV background is assumed to follow
the model by \cite{Haardt12}. This simulation fits well both the observational points
from this work and those reported by  \cite{Bolton14}(blue circles).
If we assume a slope of the TDR as in the \cite{Puchwein2015} simulation,  (i.e. the black line in the lower
panel), we can extrapolate the
curvature measurements of $T(\Delta)$ \cite[BB11,][]{Boera2014} to  mean density. The corresponding 
values of $T_0$ are shown as the dark blue connected circles
for BB11, and  the orange connected triangles for an updated version of the 
results of \cite{Boera2014}(Boera, private communication). 
The good agreement of our measurements of  $T_0$ and the evolution of $T_0$ inferred from the 
extrapolation of the temperature measurements with the curvature method  to mean density 
based on the $\gamma$ values predicted by the \cite{Puchwein2015} simulation 
is a non-trivial result that  suggests that our understanding of the thermal state IGM is converging from 
multiple  different approaches.  Convergence of results is also suggested  by the measurements of $T_0$ 
obtained with a wavelet-analysis technique by \cite{Lidz09} (cyan triangles) and 
\cite{Garzilli2012} (grey pentagons).  The discrepancy  with 
their measurements in the redshift bins close to that of our measurement
is less than  $1\sigma$.

In the lower panel, the measurement of $\gamma$ at $z=2.9$ derived from a joint 
analysis of the flux PDF and the wavelet coefficients \citep{Garzilli2012} are  clearly in  tension with our work presented here and the analysis of
\cite{Bolton14}. As discussed earlier, other measurements based on the flux PDF 
\cite[e.g.][]{VielBolton09,Calura2012} also have claimed that  an "inverted"
TDR ($\gamma <1$) is required to match the data
\citep[but see][]{Lee2012,Rollinde2013,KGpdf15}. This apparent discrepancy has 
been recognized for some time and  was recently discussed in considerable detail by \cite{Rorai2017}. By analysing a high 
signal-to-noise quasar spectrum at $z=3$, \cite{Rorai2017} confirmed  that the  
flux PDF constrains the TDR to be flat or to rise towards low density; however, as 
they point out, this is true only for gas around and below the mean density,
to which the PDF statistic is most sensitive.
\cite{Rorai2017} further show that techniques for measuring the IGM temperature that  rely on quantifying 
the smoothness of the  absorption profile, such as those discussed in this paper or the curvature method developed 
by B11, are  mainly sensitive to overdense gas, \review{i.e. 
$\Delta > 1$. This has also been highlighted in \cite{Bolton14},
where they show the relation between column density and optical
depth-weighted density (see Fig.1 in their paper)}. 
The fact that our work presented here constrains a spatially invariant TDR 
to have $\gamma > 1$ 
corroborates the conclusion of \cite{Rorai2017}  that a single, spatially-invariant
power law is not 
able to describe the TDR of the low density IGM.
This may perhaps be explained by  simulations of  HeII reionization 
where radiative transfer effects have been implemented \citep{Abel99,Paschos2007,McQuinn09,Compostella2013,LaPlante2016}, 
in which the lower densities show a bimodal temperature distribution with increased temperatures and
a flattening of the  relation between temperature and density in regions where helium has most recently
become doubly ionized.

Finally, we note that in a recent work \cite{Garzilli2015} found that the 
cutoff of the line distribution is significantly sensitive to the pressure 
smoothing, different from what our analysis shows (see Fig.~\ref{Fig:contours}).
This could be possibly due to the differences in the fitting algorithms, both 
for the individual \mlya\ lines and for the cutoff, but it is most likely related
to the particular range of column density we have selected. Fig.~8 in their paper
suggests that the effect of pressure smoothing on the low-$b$ lines is most 
prominent for lines with low column densities, while here we have only considered
those with $N_{HI}>10^{12.5}$ cm$^{-2}$, for which the line width is dominated 
by thermal broadening, as argued also by \cite{Bolton14}. 

\section{Conclusions}
\label{Sec:conclusions}
We have analysed the  \mlya\ forest of a sample of 13 
high-resolution quasar spectra in the redshift range 
$2.55 < z < 2.95$ with the help of high resolution numerical simulations of the 
IGM. The continuum-normalized 
spectra were decomposed into individual 
HI absorbers using {\sc vpfit}, and we have carefully identified and
excluded regions potentially contaminated by
metal lines.  We have then used the lower cut-off 
in the $b-N_{HI}$ distribution  and a newly 
introduced statistic based on the 
medians of the line-width distribution in separate 
column density ranges to obtain two new measurements of the temperature 
of the IGM.  In both cases we employed 
Bayesian MCMC techniques to constrain  
thermal parameters, using a grid of thermal models 
where the TDR has been imposed in post processing. 
Our results can be summarized as follows.
\begin{itemize}
\item Fitting the lower cut-off of the $b-N$
distribution  in the standard way gives  $T_0/ 10^3\, {\rm K}=14.3 \pm 5.0$ K
and $\gamma=1.52\pm 0.22$. 


\item The parameters describing the lower cut-off 
and  that describing  the TDR are strongly correlated in simulations, but   
there is significant scatter due to   the residual dependence on other
thermal parameters (in particular  the smoothing length $\lambda_P$).

\item We have introduced a new  calibration of the temperature measurement 
based  on a  linear combination of the thermal parameters (in logarithmic space)
whose coefficients are left free to vary.  For this new calibration we have implemented
a  MCMC analysis in which the calibration coefficients are  estimated  together with the 
thermal quantities $T_0,\gamma$ and $\lambda_P$. 

\item Including  the correlation between the statistical
uncertainties of $\log b_0$ and $\Gamma$ in our likelihood analysis
gives   $T_0/ 10^3\, {\rm K}=15.6\pm 4.4$ K and $\gamma= 1.45 \pm 0.17$. Taking the correlation
between $T_0$ and $\gamma$ shows that  the inferred values for the two parameters are 
strongly degenerate. Conversely, no sensitivity 
is found to the smoothing length $\lambda_P$.

\item We have introduced an alternative  statistical estimator  for the 
thermal parameters which is based on the medians $m_1$ and $m_2$ 
of the line-width distribution for  $N < 10^{13.5}$ cm$^{-2}$
and $N > 10^{13.5}$ cm$^{-2}$, respectively. For this we have defined the 
transformation $\log b'=\log b + \alpha (\log N_{HI} - 13)$,
with $\alpha$ chosen such that $m_1=m_2$ for $b'$. 

\item By applying the parameters $m_1$ and $\alpha$ to the same
MCMC technique used for the lower cut-off method, we obtain
$T_0/ 10^3\, {\rm K}=14.6 \pm 3.7$ K and $\gamma=1.37 \pm 0.17$, in good agreement 
with our measurement from the lower cut-off.  For the measurement 
based on the differential median method 
we find a strong degeneracy with the smoothing length $\lambda_P$,
suggesting, unsurprisingly, that the overall $b-N$ distributions  contains information 
about  the small-scale spatial structure of the IGM.  

\item When we use the posterior distribution for the TDR parameters to
infer the temperature at $\Delta \approx 3$, we obtain  values consistent
with the measurements at the same redshift from BB11
and \cite{Boera2014}.

\item Our measurements are  also in good agreement with the 
theoretical predictions from the hydrodynamics model of
\cite{Puchwein2015} and with the measurements of $T_0$ at 
similar redshifts from \cite{Lidz09} and \cite{Garzilli2012}.

\item Our constraints on $\gamma$ are in disagreement with 
claims of an inverted TDR based on the flux PDF, similar  
to published results using  statistics based on the  
smoothness of the absorption. Our findings  further corroborate 
those of \cite{Rorai2017}, who argued that this is due
to the different overdensity range probed by the PDF. 
\end{itemize}

The work presented here marks a further step toward a consistent 
characterization of the thermal state of the IGM at $z \lesssim 3$. 
The agreement with other recent measurements at the same redshift and with
high resolution hydrodynamics models is an encouraging sign that
our  understanding  of the thermal state of the IGM is converging. 
The techniques developed and presented here can be easily
applied to datasets at other redshifts and should improve the well established
approach of characterising the thermal state of the IGM with help of the 
$b-N$ distribution of \mlya\ forest absorbers.  Our analysis 
takes into account the uncertainties in the calibration between
the lower cutoff (or the median $b$-parameters in different column density ranges) and the thermal parameters, as well as 
parameter correlations and second-order dependencies which were previously neglected. 
The degeneracies we find, especially in our  analysis based on the  differential median, suggest that a joint analysis with  different 
statistics constraining the pressure smoothing length will significantly  improve the precision of  current constraints 
on the thermal state of the IGM.

\section*{Acknowledgements}
We are grateful to W.L.W.Sargent for kindly providing two of the spectra analysed in
this work, Volker Springel for making Gadget-3 available and Girish Kulkarni 
for sharing the code for the calculation of $\lambda_P$ in SPH simulations. \review{We thank the anonymous referee for 
useful comments and suggestions, which greatly improved
this manuscript.}
MH acknowledges support by ERC ADVANCED
GRANT 320596 ``The Emergence of Structure during the epoch of Reionization''. GB was supported by the National Science Foundation through grant AST-1615814. JSB 
acknowledges the support of a Royal Society University Research
Fellowship. MTM thanks the Australian Research Council
for \textsl{Discovery Project} grant DP130100568.  This work
made use of the DiRAC High Performance Computing System (HPCS) and the
COSMOS shared memory service at the University of Cambridge. These are
operated on behalf of the STFC DiRAC HPC facility. This equipment is
funded by BIS National E-infrastructure capital grant ST/J005673/1 and
STFC grants ST/H008586/1, ST/K00333X/1.





\bibliographystyle{mnras}
\bibliography{line_fitting}
\appendix

\section{Effect of Temperature and Optical Depth Variation}
\label{appendix}

\begin{figure*}
\includegraphics[width=\textwidth]{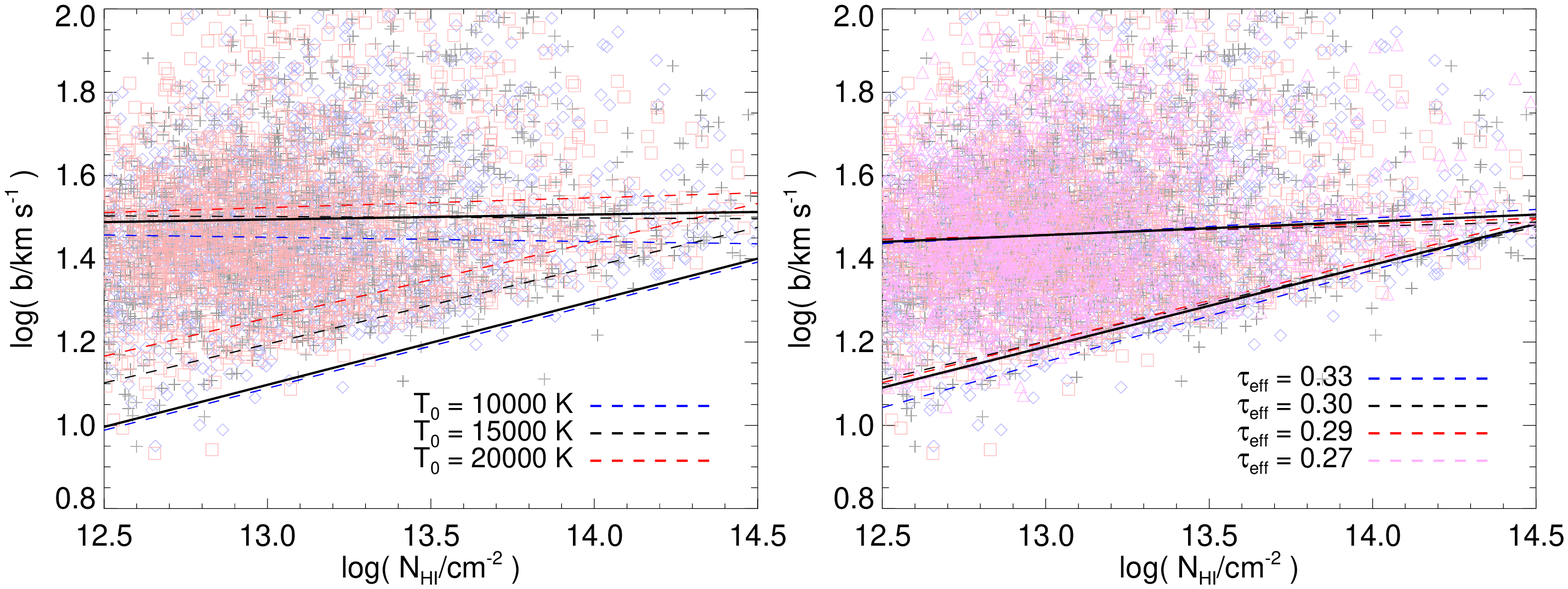}
\caption{\label{fig:mixture_test} Effect of mixed temperatures
or effective optical depths on the cut off and on the differential
median. \emph{Left panel:} a combination of three models with 
$T_0 = 10000 $ K (blue diamonds and lines), 
$T_0 = 15000 $ K (black crosses and lines)
and $T_0=20000$ K (red squares and lines). The differential median is represented as in Fig.~\ref{Fig:median_statistic}, by dashed lines
for the individual models and by a thick solid black line for the
combination of all of them. The same is done for the lower cut-off.
\emph{Right panel:} same as the left panel, but for a combination
of four models with different  effective optical depth 
$\tau_{\rm}=0.33$ (blue)
$\tau_{\rm}=0.30$ (blue)
$\tau_{\rm}=0.29$ (blue)
$\tau_{\rm}=0.27$ (blue). }
\end{figure*}

\review{To understand the effect of an evolving optical depth and a possible variation in temperature, we conducted the following test: 
\begin{itemize}
\item  we created a model which is a mixture of four models with different effective optical depth, $ \tau_{\rm eff}=-\log \bar{F} = 0.33, 0.3,  0.29,  0.27 $. This values encompass the evolution of the mean transmission across our redshift bin. 
\item we created another model which is a mixture of three models with different temperatures at mean density $T_0 = 10000, 15000, 20000$ K. This mixture could be interpreted either as a (rather  extreme) temperature evolution with redshift or as spatial fluctuations.
\end{itemize}
We then calculated the cut-off and the differential median statistic for the two mixed models and for their individual components. The results are shown in Fig.~\ref{fig:mixture_test}.  }

\review{In the case of the optical depth mixture (right panel), the differences between the total model (black solid lines) and the four components (coloured dashed lines) are minimal, especially for the differential median statistic. 
One of the four cutoffs slightly deviates from the others at low column densities, but given that there is no clear trend and given the magnitude of the error (the outlier  model has an uncertainty of $\Delta \log b_0 = 0.065$, four times larger than the other models), we consider it a statistical fluctuation due to the noise/bootstrap realizations. }

\review{In the case of the temperature mixture (left panel), it appears evident that the differential median of the total model (black solid) depends on the average temperature, i.e. coincides with the middle of the three components, while the cut-off picks the coldest of the three. This suggests that, if there is significant temperature evolution or fluctuation, measurements based on the cut-off will be biased towards the lower values, differently from the median statistic. However, our measurements based on the two methods are in good agreement, and if anything the cut-off analysis points toward slightly higher temperatures, which leads us to conclude that this effect is negligible at this level of precision. 
}


\bsp	
\label{lastpage}
\end{document}